\documentclass{arxiv_old}
\usepackage[normalem]{ulem}

\usepackage{float}
\pdfoutput=1

\usepackage{amssymb,amsfonts,amsmath,graphics,graphicx}

\begin{document}






\title{Alienation in italian cities. Social network fragmentation from collective data.}

\author{Pierluigi Contucci\affil{2}{Dipartimento di Matematica, Universit\`a di Bologna, Italy},
Cecilia Vernia\affil{3}{Dipartimento di Scienze Fisiche, Informatiche e Matematiche, Universit\`a di Modena e Reggio Emilia, Italy}}

\contributor{corresponding author: cecilia.vernia@unimore.it}

\maketitle
%
%

\begin{article}
\begin{abstract}


We study the structure of a social network of strong ties (trust network) investigating its property of connectedness versus fragmentation. To this purpose we analyse an extensive set of census data, about marrying or having children with immigrants, collected by Italian national statistical institute for all Italian municipalities from 2001 to 2011. Not using neither obtaining personal local information but only average ones, our method fully complies with privacy and confidentiality. Our findings show that large cities display the behaviour of highly fragmented 
trust networks where individuals face possible phenomena of alienation. Smaller cities and villages 
instead behave like fully connected social systems with a rich tie structure, 
where isolation is rare or completely absent. While confirming classical 
sociological theories on alienation in large urban areas our approach 
provides a quantitative method to test them and a predictive tool for policy makers.

\end{abstract}

Human choices, as clearly elucidated in the work of Max Weber \cite{weber}, can be
distinguished according to the fact that they are {\it social actions} or not. Sometimes, mostly when
the matter of the choice is simple and the conclusions can be reached with elementary
reasoning, we draw a decision by ourselves, without relying on other opinions and
only counting on our pre-existing cultural background. Some other times (the case of social action), 
when the choice is of complex nature or when the information we have is not enough to draw a rational
conclusion we proceed, {\it whenever possible}, by relating to others either by imitation or seeking
for interaction and consensus. Those two types of individual choice mechanisms 
lead to two different emerging collective social behaviour each displaying 
peculiar features. The case of independent choices usually come with 
smooth, locally linear, behaviour while the other 
may present different characteristics, including tipping points, 
according to the nature and structure of the interactions among the
individuals and the topological features of the underlying social network.

In this paper we investigate, by analysing an extensive dataset about Italy,  
some features of the {\it trust} social network \cite{trust, trust2, pesco, tilly} involved with two specific choices.
These are the choice of marrying a person from your own country or from a different one,
and the choice of having a child under the same two alternatives.
Those {\it coupling} choices are among the most important people face in their existence. This is due not 
only to their strong intrinsic emotional content but also because norms
and regulations, of both moral and legal nature, make them basically life long.

\begin{figure}[t]
          \centering
               \includegraphics[width=10cm,height=10cm]{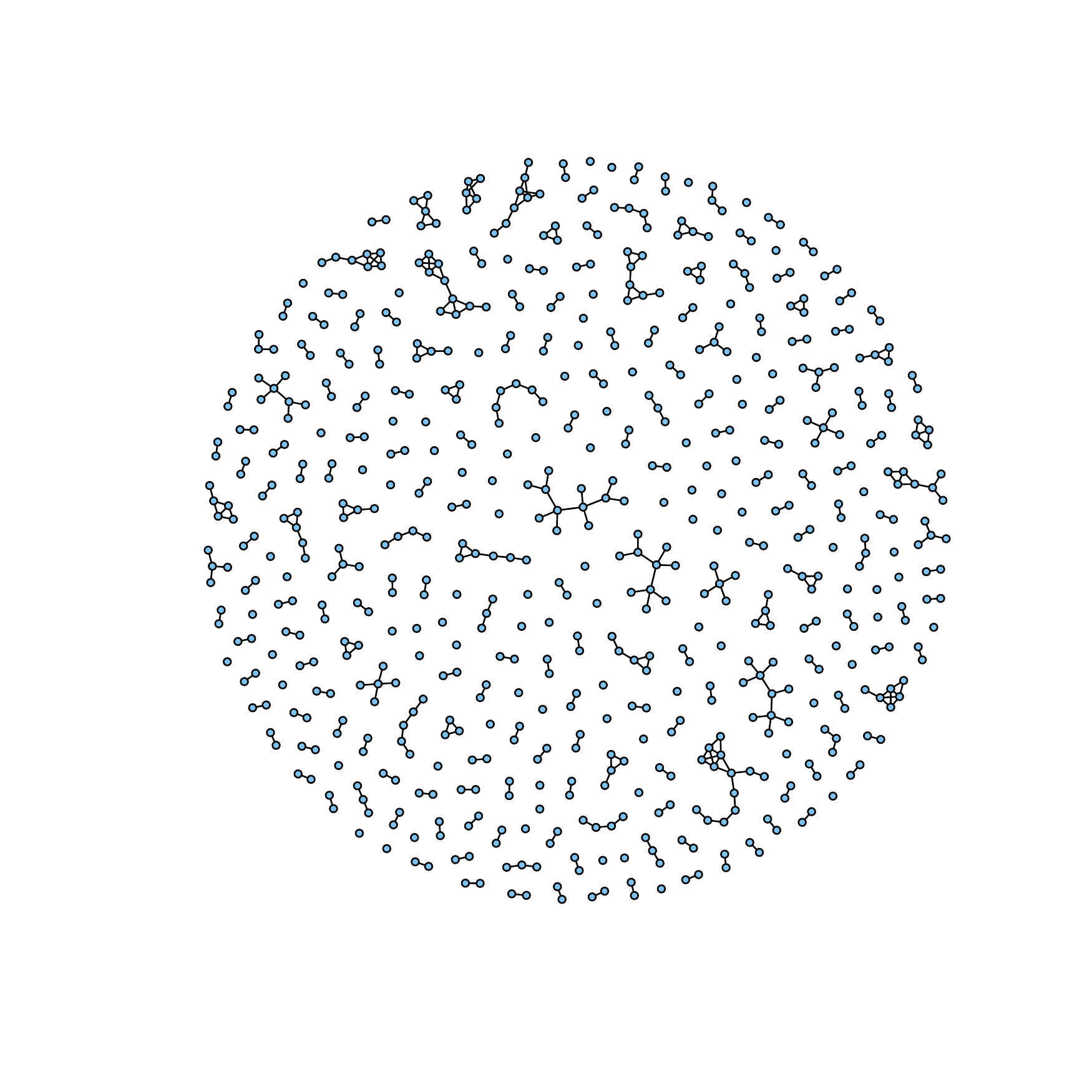}
\caption{\label{FigUnp} Local view of a sparse unpercolated graph. Individuals are either isolated or, seldomly, connected with few others.
Our findings indicate this structure for the average local trust network in large Italian cities}
\end{figure}

\begin{figure}[t]
          \centering
               \includegraphics[width=10cm,height=10cm]{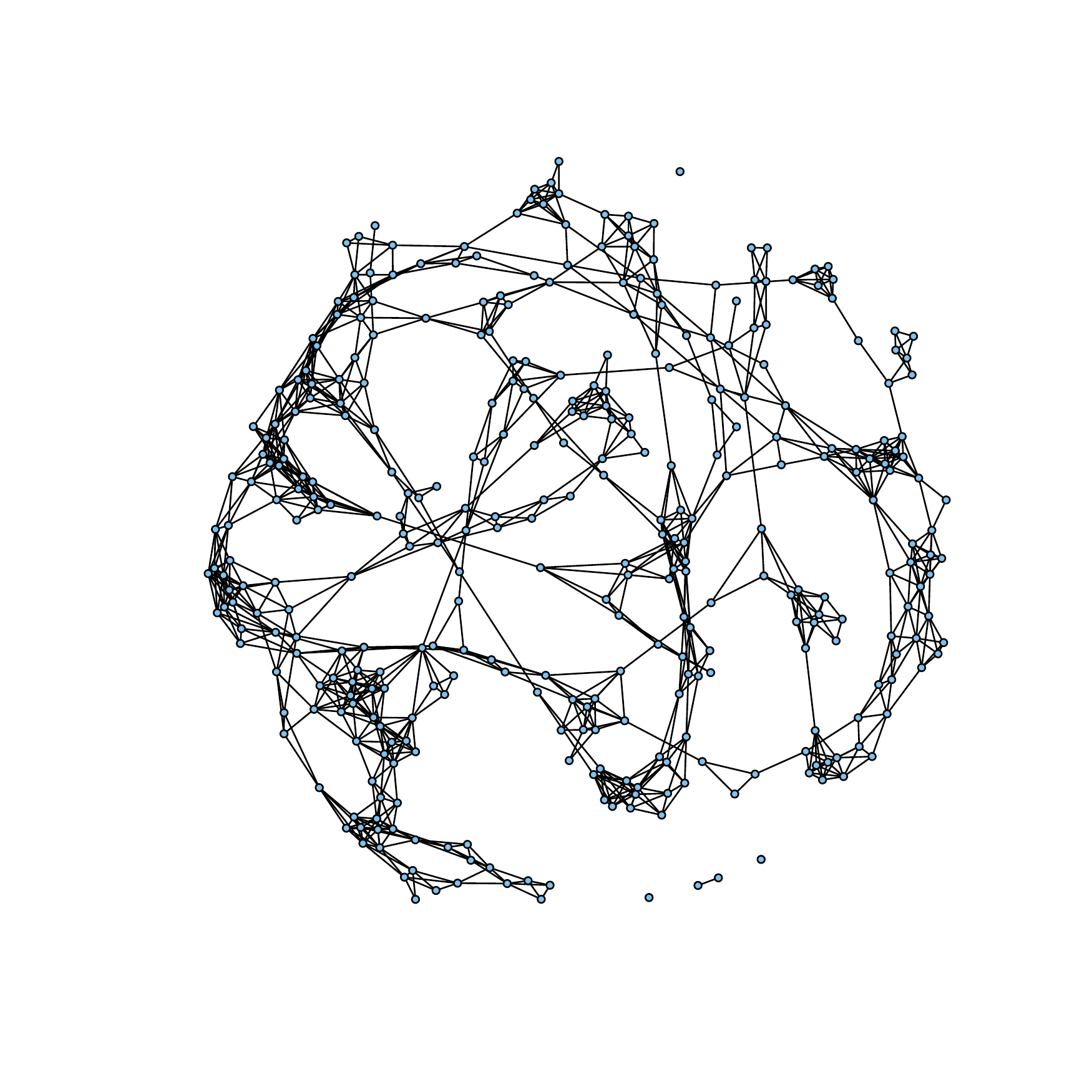}
\caption{\label{FigPerc} Local view of a percolated graph. Individuals are connected with several others. The graph, a part for few individuals, is completely connected. Our finding indicate this structure for the local average trust network in small Italian cities.}
\end{figure}

Such choices, with the complexity they come with, fully belong to the category of social actions.
This is both a sociological finding \cite{bbf,bbs} and also a recent conclusion
of the work \cite{bcsv}. Possible differences observed in their collective social appearance depend 
therefore on the underlying social network. In particular if the social network is 
extremely sparse (Figure \ref{FigUnp}), with links that are so rare that do not allow the whole group to connect
and percolate, we expect a social behaviour similar to a group of independent
individuals. Conversely, if the network is globally well connected (Figure \ref{FigPerc}), with rare cases of isolated
individuals or small groups, we expect a genuine collective effect to emerge.

Distinguishing among these two types of social network has important sociological
implications. The network involved with the two mentioned choices is indeed
a crucial one since it is made by strong personal ties \cite{ties, grano} possibly affecting
also other relevant personal choices. Being part of a well connected social group,
with a shared and mutually approved behaviour, provides individuals with a sense of moral rightness, 
safety and availability of support. If instead the network is sparse and the connected groups 
are made of few units the same perception of belonging to a group is lacking and 
individuals feel alienated \cite{segre, durkheim}. 

The key idea that we advance in this paper is that it is possible, from the the observation
of the global social behaviour, to infer what type of network the choice is holding on.
In order to derive the network structure we investigate how the frequency of each different
coupling choice changes when immigrant's percentage increases. In particular we analyse Italian
data about the frequency of mixed marriages and 
newborn children from mixed couples. Our findings show that, up to an immigrant density
 of 7\%, there are two different growth laws according to the size of the municipality. 
For large cities (with more than 10K inhabitants) if the frequency has grown to a 
given amount $\Phi$ when the immigrant density is 1\% it will grow to 2$\Phi$ at 2\%, 3$\Phi$ at 3\% and so on. 
In small cities the behaviour is different: in order to reach the frequency 2$\Phi$ the immigrant density has to grow to 4\%, 
to reach 3$\Phi$ it has to grow to 9\% and so on. While the first law is linear, i.e. has a constant growth 
rate the second, that follows the square root function, has an anomalous high growth rate 
when the phenomenon starts, which quickly decreases at increasing immigrant densities.

We interpret these results within the statistical physics framework which is largely used in 
many research fields to study the behavior of quantities that can be represented as large 
sums of mutually correlated random variables (see \cite{bcsv} and references there in). 
On the basis of a statistical physics model 
developed in \cite{bcsv} we argue that the two different growth laws
depend on the different social network structures. In particular the large italian cities display the behavior 
of sparse unpercolated networks while the small ones behaves like fully connected networks. 
In other words individuals living in large italian cities turn out to be in alienated conditions
while smaller cities provide a full social environment. This conclusion provides a 
quantitative confirmation of the general sociological theories about alienation and 
anomie in large cities \cite{durkheim}. 

Our study doesn't use nor produce any information of proximity
like who trusts who and is fully compliant with privacy regulations. Starting from collective 
information we deduce average topological properties of the network. Moreover the 
fact that our approach is purely of observational nature and doesn't present any type 
of solicitation to individuals nor to the system as a whole, makes our approach ethically
viable. 

\section{Data description and results}

We consider here a database  
on mixed marriages and newborns to mixed couples, annually 
recorded by ISTAT (Italian Institute of Statistics) for all italian municipalities 
in the time period from 2001 to 2011.
In that time span documented foreign born population has grown in Italy  from just over 1 million to almost 4 millions people, corresponding to an increase from $2\%$ to over $6\%$ of the total resident population. These figures do not include illegal immigrants whose numbers are difficult to determine. Although there are immigrants from almost all nations, the communities with higher presence are from Romania ($969$,$000$ in $2011$, i.e $21\%$ of the total population of immigrants), followed by Albania ($483$,$000$ in $2011$, i.e $11\%$), Morocco ($452$,$000$ in $2011$, i.e $10\%$), and China ($210$,$000$ in $2011$, i.e $5\%$). The geographical distribution of  immigrants is largely not homogeneous:
$86.5\%$ lives in the northern and central part of the country (the most economically developed areas), while $13.5\%$ lives in the south.

The dataset contains over $1$,$100$,$000$ data,
yearly describing - for each of the $8$,$100$ municipalities - the total population,
the number of immigrants, the number of marriages and newborns 
originating from different types of couples (either mixed or not).
For each municipality we have considered the immigrant density
\begin{equation}
\centering\gamma=\frac{N_{imm}}{N_{imm}+N_{nat}} \; ,
\end{equation}
where $N_{imm}$ is the number of immigrants and $N_{nat}$ is the number of the natives.
A useful parameter, measuring the number of possible cross-links among natives and
immigrants is $\Gamma=\gamma(1-\gamma)$.
We focussed, in particular, on two quantifiers: the fraction 
of marriages with spouses of mixed origin (native and immigrant) $M_m$ and the fraction of newborns 
with parents of mixed origin $B_m$.
The entire marriages dataset contains $89$,$093$ records but only 
$82$,$208$ points were considered in our study.
This is due to the fact 
that in tiny villages, occasionally, no marriages occurred. Those events
account for about $8\%$ of the whole dataset. Moreover, in about 
$56\%$ of all records no mixed marriages have occurred. 
Analogously the inputs 
from municipalities where no newborns occurred account for 
about $4\%$ of the whole dataset, whereas for about $43\%$ 
of all records no mixed newborns have occurred. 

\begin{figure}[h!]
          \centering
               \includegraphics[width=8.5cm,height=6cm]{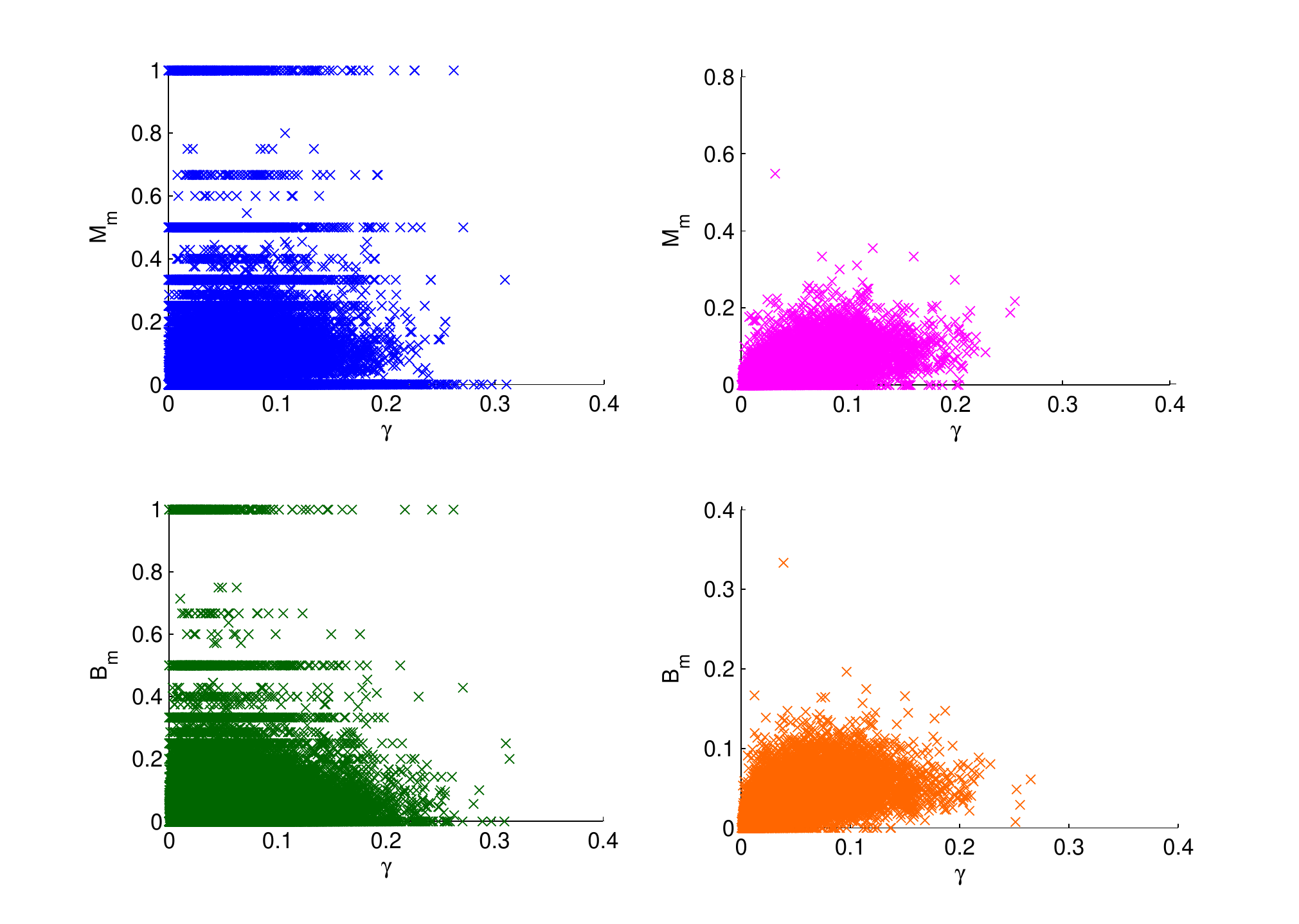}
\caption{\label{FigRaw} Raw data versus immigrant density $\gamma$. Blue points represent the fraction of mixed marriages occurred in municipalities with less than $10$,$000$ inhabitants where a percentage $\gamma$ of migrants is present; similarly green points account for newborns from mixed couples. Magenta points represent the fraction of mixed marriages occurred in municipalities with more than $10$,$000$ inhabitants, while orange ones mirror the newborns from mixed couples. One may note that data in the left panels seem to lie along horizontal lines displaced according to $1/n$, with $n\in\mathbb{N}$ due to effects seen in small-sized municipalities.}
\end{figure}

\begin{figure}[h!]
          \centering
               \includegraphics[width=8.5cm,height=6cm]{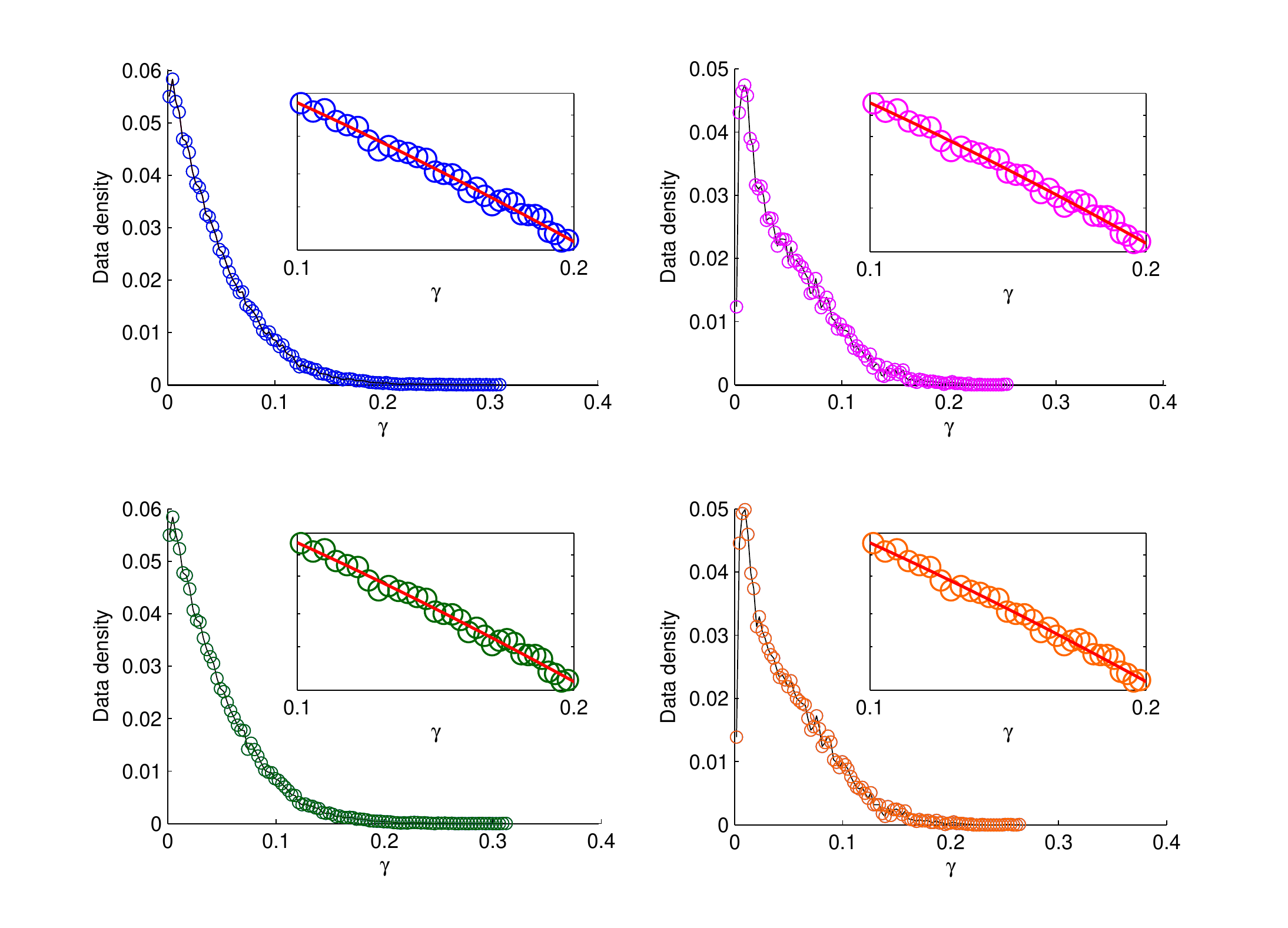}
\caption{\label{FigDens} Data density for the mixed marriages and mixed newborns quantifiers as a function of the percentage of immigrants $\gamma$. Left upper panel: density of the mixed marriages (blue circles) dataset for municipalities with less than $10$,$000$ inhabitants. Right upper panel: density of the mixed marriages (magenta circles) dataset for municipalities with more than $10$,$000$ inhabitants. Lower panels display the same information for the newborns in small (green circles) and large (orange circles) municipalities. In the insets, the marriages and newborns data density plots are fitted, for $\gamma\geq0.1$, with the power-law behavior (in log-log scale) where $\mu(\gamma)\propto\gamma^\delta$ with $\delta=-4.6\pm0.24$ (small municipalities) and $\delta=-5.4\pm0.31$ (large municipalities). In all panels the data densities decrease close to $\gamma=0$. 
This comes from the fact that
the migration phenomena was already ongoing when the data collection was started ($2001$) and the density 
of migrants in Italy was larger than zero.}
\end{figure}

We separated the dataset into two parts: one for
the small cities (below $10$,$000$ inhabitants) and the other for large ones (above $10$,$000$ inhabitants).
Figure \ref{FigRaw} shows the resulting collection of the raw data of each part for the mixed marriages (upper panels) 
and for the newborns of mixed couples (lower panels) in the planes $(\gamma,M_m)$ and $(\gamma,B_m)$, respectively. 
We observe that records belonging to villages and small cities,
whose population is under 
$10$,$000$ inhabitants, count up to $84\%$.\
Figure \ref{FigDens} displays the data densities
and shows statistical robustness (up to one percentile) for all $\gamma$ up to $16\%$.

The partition of both marriages and newborns datasets, which has led 
us to the main results of this work, has been investigated under various 
perspectives. We remark that this partition was proposed 
since an analysis performed over a unified datasets (small and large cities together) 
happened to be unsatisfactory. It displayed in fact a high dependence on 
the binning parameter settings revealing the typical presence
of {\it data mixture} of interacting and non-interacting type. 
This situation led us to partition each dataset, separating 
small municipalities from large ones. The implemented threshold 
(i.e. $10$,$000$ inhabitants) was attempted according to previous work 
\cite{bcsv}, where the considered municipalities consisted only of cities 
over $10$,$000$ inhabitants. Ex-post an optimisation test on the threshold has been 
performed showing that the proposed one was a stable 
choice, i.e. the coefficients of determination ($R^2$) of the data fitting in  the two regimes were maximal.
Moreover, for the chosen partition we verify both robustness and homogeneity 
among the two datasets (large and small cities) in order to exclude
pathologies by imbalance. Our tests show that the immigrant proportions over the total population in large and small cities are comparable for each year
with a maximum difference between the two of about one percent. We can therefore infer that immigrants seem to be distributed in the same manner, independently from the size of the municipality they belong to. 
Eventually, 
the ratio between the total population living in small municipalities and the one living large municipalities 
is nearly constant ($\sim 47\%$) over the years.
We tested, moreover, the behavior of the temporal series (year by year) of the proportions of immigrants, of  natives 
and of the total population living
in the small cities. This study shows that 
 the national fraction of immigrants and natives in small municipalities are about the same ($30\%$).
Furthermore, a similar value (slightly more than $30\%$) is found for the total population that lives in cities with less than $10$,$000$ inhabitants.

\begin{figure}[h!]
          \centering
               \includegraphics[width=8.5cm,height=6cm]{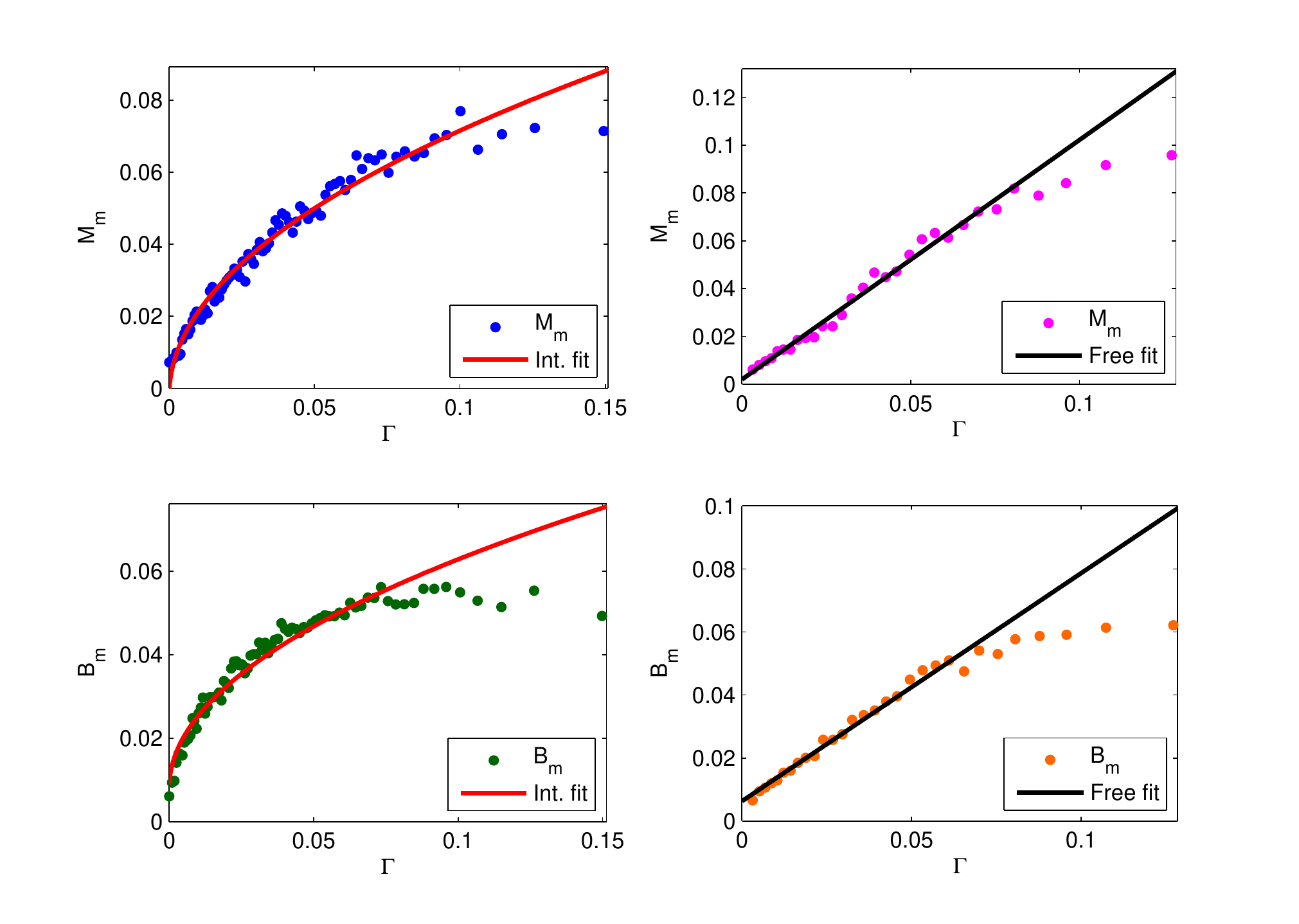}
\caption{\label{FigRadLine} Dots are average quantities for the the mixed marriages and mixed newborns quantifiers versus $\Gamma$. Left upper panel: quantifier $M_m$ (blue dots), fraction of mixed marriages occurred in municipalities with less than $10$,$000$ inhabitants, with the best square root fit (red curve) $a\sqrt\Gamma+b$ (a = 0.233 $\pm$0.009, b = -0.002$\pm$0.002, with a goodness of fit $ R^2$ = 0.97 computed for $\Gamma$\textless0.13). Right upper panel: quantifier $M_m$ (magenta dots), fraction of mixed marriages occurred in municipalities with more than $10$,$000$ inhabitants, with the best linear fit (black curve) $a\Gamma+b$ (a = 1.00 $\pm$0.05, b = 0.002$\pm$0.002, with a goodness of fit $R^2$ = 0.98 computed for $\Gamma$\textless0.08). 
Left lower panel: quantifier $B_m$ (green dots), fraction of newborns with mixed parents, born in municipalities with less than 10.000 inhabitants, with the best square root fit (red curve) $a\sqrt\Gamma+b$ (a = 0.174 $\pm$0.008, b = 0.008$\pm$0.002, with a goodness of fit $ R^2$ = 0.96 computed for $\Gamma$\textless0.10). Right lower panel: quantifier $B_m$ (orange dots), fraction of newborns with mixed parents, born in municipalities with more than $10$,$000$ inhabitants, with the best linear root fit (black curve) $a\Gamma+b$ (a = 0.78 $\pm$0.04, b = 0.005$\pm$0.001, with a goodness of fit $ R^2$ = 0.99 computed for $\Gamma$\textless0.07)}
\end{figure}

Since our study is devoted to investigate the average behavior at the country scale for the two quantifiers $M_m$ and $B_m$ in large cities and small ones, we estimate the average percentage of mixed marriages or mixed newborns, for a given 
immigrant density. To this purpose we performed a {\it mediant} average and binning with constant information (see \cite{bcsv} for details).
Figure \ref{FigRadLine} displays the output averages versus $\Gamma$. The emerging behaviors are well fitted (see the $R^2$ coefficient) by two
different laws: square root for the quantifiers on small municipalities and linear on large ones.

The statistical physics approach developed and broadly described in the technical paper \cite{bcsv} suggests an interpretation for these results. 
We know in fact that the linear behaviour emerges, in strong ties conditions, from collective effects when the network is sparse and unpercolated. 
The links are rare and the connected groups are made of only few units. On the other side, the square root law emerges when the social framework 
of strong personal ties, is built on a fully connected network where the cases of isolated individuals or small groups inside the communities are rare
or completely absent.

\begin{figure}[h!]
          \centering
               \includegraphics[width=8.5cm,height=6cm]{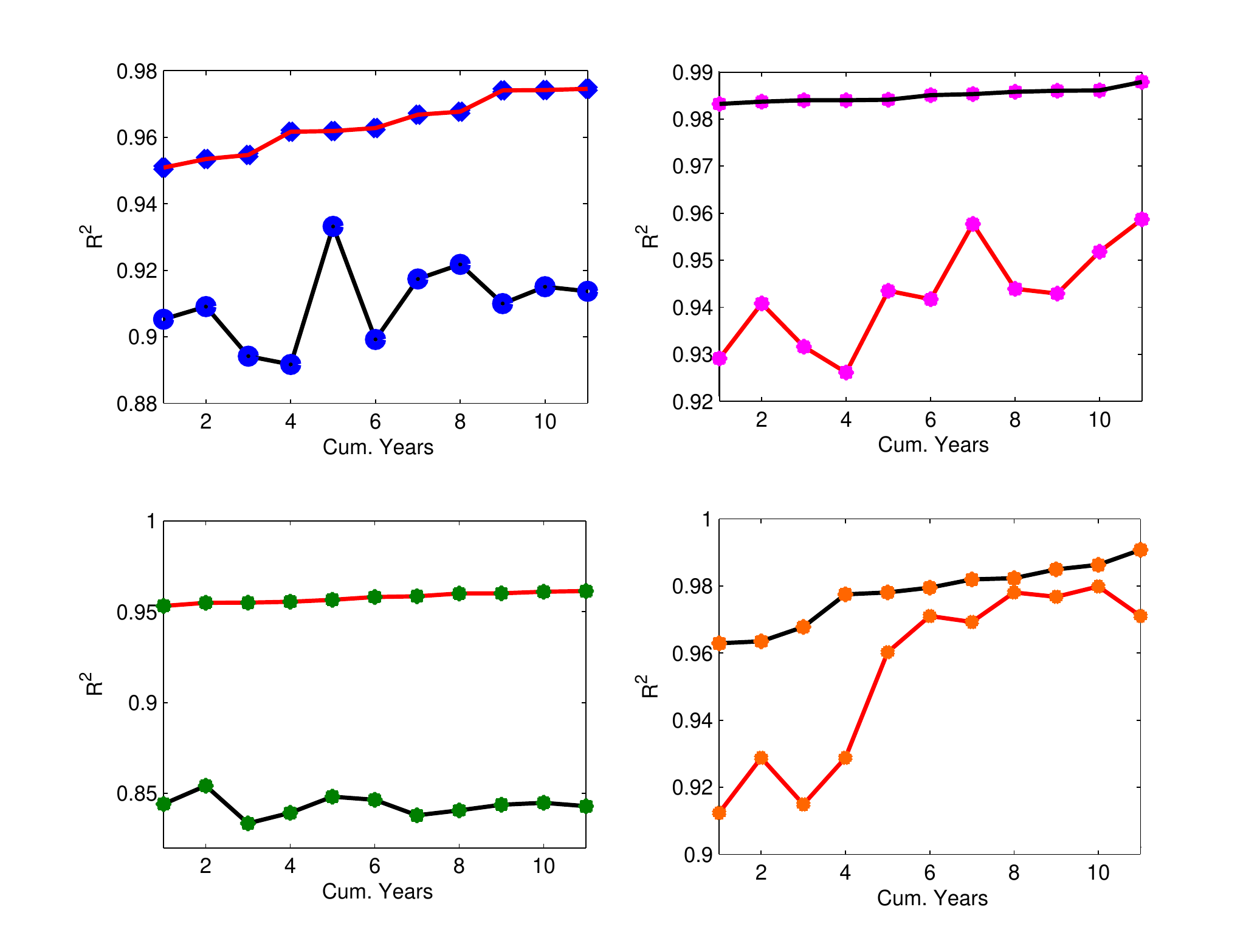}
\caption{\label{FigR2} Determination coefficients $R^2$ of different fits (linear and square root functions) for the averaged quantifiers shown in figure 3
against an increasing cumulative time span: the first cumulative time span includes data from 2001, the second cumulative time span
data from the interval 2001-2002, up to the last that includes data from the entire period $2001-2011$. $R^2$ are represented through red curves when related to a square root fit and through a black curve when related to a linear fit. Left upper panel: $R^2$ (blue dots) for the mixed marriages occurred in municipalities with less than $10$,$000$ inhabitants. Right upper panel: $R^2$ (magenta dots) related to mixed marriages occurred in municipalities with more than $10$,$000$ inhabitants. Analogously in the lower panels for the newborns in small (green circles) and large (orange circles) municipalities.}
\end{figure}

In order to better test the predictability power of our theory we have performed our analysis not only for whole time span
of the observation data, but also on increasing time sub-spans. Figure \ref{FigR2} shows that the distinction among 
linear and square root growth is already evident from elaboration coming from only $2001$ data, and stay
stable for increasing intervals $2001-2002$ etc, up to the whole set of data ($2001-2011$).

Our results are in good agreement with the classical sociological theories of alienation and anomie about the social behavior on large
cities, where social connections are seldom and ineffective, in comparison to villages where
they are strong (see \cite{durkheim}). 
Our results moreover have the potential to impact policy makers. In fact, the identification of the growth law of each quantifier
in $\gamma$ allows to predict the level of the quantifiers for growing values of $\gamma$.


\begin{acknowledgments}
We thank Adriano Barra and Rickard Sandell for interesting discussions. We also acknowledge Claudio Giberti for valuable suggestions and for a careful reading of the manuscript. We thank Francesco de Pretis for his help at a very early stage of the manuscript.
This work was supported by the FIRB grant RBFR10N90W and by the PRIN grant 2010HXAW77.
\end{acknowledgments}

\begin{authorinfo} 
Correspondence and requests for materials should be
addressed to {\bf corresponding author}.
\end{authorinfo}

\vfill\eject
\newpage

\vfill\eject

\end{article}


\begin{thebibliography}{}

\bibitem{weber} Weber, M., {\em Economy and Society: An outline of interpretative sociology}, [p.23], 
 (California University Press, 1978).

\bibitem{trust} Bravo, G., Squazzoni, F., Boero R., Trust and partner selection in social networks: An experimentally grounded model, {\em Social Networks} \textbf{34}, 481-492 (2012).

\bibitem{trust2} Sherchan, W., Nepal, S., Paris, C.,  A Survey of Trust in Social Networks, {\em ACM Comput. Surv.}, \textbf{45}, 47:1-47:33 (2013).

\bibitem{pesco} Pescosolido, B.A., The Sociology of Social Networks,  {The Handbook of 21st Century Sociology}, 208-217, (C.D Bryant and D.L. Peck, eds. Thousand Oaks, CA, Sage Publications 2006). 

\bibitem{tilly} Tilly, C., {\em Trust and Rules}, (Cambridge University Press New York, 2005).

\bibitem{bbf} Blau, P.M., Beeker, C., Fitzpatrick K.M., Intersecting Social Affiliations and Intermarriage
{\em Social Forces} \textbf{62}, 585-606 (1984).

\bibitem{bbs} Blau, P.M., Blum, T.C., Schwartz,J.E., Heterogeneity and Intermarriage {\em American Sociological Review} \textbf{47}, 45-62 (1982)


\bibitem{bcsv} Barra, A., Contucci, P., Sandell, R., Vernia, C., An analysis of a large dataset on immigrant integration in Spain. The Statistical Mechanics perspective on Social Action.
{\em Sci. Rep.} \textbf{4}, 4174, (2014) .

\bibitem{ties} Easley, D. \& Kleinberg, J., {\em Networks, Crowds, and Markets: Reasoning about a Highly Connected World.} [p. 47] (Cambridge University Press, 2010).

\bibitem{grano} Granovetter, M.S., The strength of weak ties,  {\em Am. J. Sociol.}, \textbf{78}, 1360-80 (1973).

\bibitem{segre} Di Prete, T.A., Gelman, A., McCormick, T., Teitler, J., \& 
Zheng, T., Segregation in Social Networks Based on Acquaintanceship and Trust, {\em Am. J. Sociol.}, \textbf{116}, 1234-83 (2011).

\bibitem{durkheim} Durkheim, E.,  {\em Le Suicide: Etude de sociologie.} Félix Alcan Paris. (1897).

\end{thebibliography}
\end{document}